\documentclass[conference,10pt,letterpaper]{IEEEtran}

\usepackage{url}
\usepackage{cite}
\usepackage{graphicx}
\usepackage{hyperref}
\usepackage{svg}
\usepackage{booktabs}
\usepackage{hyperref}
\usepackage{soul}

\title{Consistent and Repeatable Testing of O-RAN Distributed Unit (O-DU) across Continents}

\author{
    \IEEEauthorblockN{Tuan V. Ngo\IEEEauthorrefmark{1}, Mao V. Ngo\IEEEauthorrefmark{1}, 
    Binbin Chen\IEEEauthorrefmark{1},  
    Gabriele Gemmi\IEEEauthorrefmark{2},
    Eduardo Baena\IEEEauthorrefmark{2},
    Michele Polese\IEEEauthorrefmark{2},\\ 
    Tommaso Melodia\IEEEauthorrefmark{2},
    William Chien\IEEEauthorrefmark{3}, 
    and Tony Quek\IEEEauthorrefmark{1} 
    }
    \IEEEauthorblockA{\IEEEauthorrefmark{1}Singapore University of Technology and Design
    \{vantuan\_ngo, vanmao\_ngo, binbin\_chen, tonyquek\}@sutd.edu.sg}
    \IEEEauthorblockA{\IEEEauthorrefmark{2}Institute for the Wireless Internet of Things, Northeastern University \{g.gemmi, e.baena, m.polese, t.melodia\}@northeastern.edu}
    \IEEEauthorblockA{\IEEEauthorrefmark{3}Synergy Design Tek
    william\_chien@synergydesigntek.com}
}

\begin{document}

\maketitle

\begin{abstract}
Open Radio Access Networks (O-RAN) are expected to revolutionize the telecommunications industry with benefits like cost reduction, vendor diversity, and improved network performance through AI optimization. Supporting the O-RAN ALLIANCE's mission to achieve more intelligent, open, virtualized and fully interoperable mobile networks, O-RAN Open Testing and Integration Centers (OTICs) play a key role in accelerating the adoption of O-RAN specifications based on rigorous testing and validation. One theme in the recent O-RAN Global PlugFest Spring 2024 focused on demonstrating consistent and repeatable Open Fronthaul testing in multiple labs. To respond to this topic, in this paper, we present a detailed analysis of the testing methodologies and results for O-RAN Distributed Unit (O-DU) in O-RAN across two OTICs. We identify key differences in testing setups, share challenges encountered, and propose best practices for achieving repeatable and consistent testing results. Our findings highlight the impact of different deployment technologies and testing environments on performance and conformance testing outcomes, providing valuable insights for future O-RAN implementations.


\end{abstract}


\section{Introduction}

The O-RAN architecture aims to create more flexible, interoperable, and intelligent 5G and future wireless networks through openness, programmability, and hardware-software disaggregation. This approach is expected to improve network deployment, operation, and maintenance while also promoting innovation and cost-efficiency~\cite{polese2023_understanding, Mao_ICT_Express24}. The O-RAN Distributed Unit (O-DU) is a critical component in the O-RAN architecture, which hosts the Radio Link Control (RLC), MAC, and High-PHY layers based on a lower-layer functional split. The conformance, performance, and reliability of O-DU, as well as its interoperability with other O-RAN components, are essential for successful O-RAN deployments.

Testing O-RAN components, particularly the O-DU, presents unique challenges due to the disaggregated nature of the architecture and the involvement of multiple vendors. The O-RAN 
ALLIANCE, through its Test and Integration Focus Group (TIFG)~\cite{o_ran_end_to_end_test_specification}, has initiated efforts in its recent Global PlugFest Spring 2024 to demonstrate consistent and repeatable testing of open fronthaul (OFH) testing in multiple labs \cite{PlugFestTheme}. 
%
Towards this,
collaborative efforts between Open Testing and Integration Centres (OTICs) have been established. Specifically, the Northeastern University Open6G OTIC in the Boston area\footnote{\url{https://wiot.northeastern.edu/otic}} and the Asia \& Pacific OTIC in Singapore (APOS)\footnote{\url{https://fcp.sutd.edu.sg/otic/}} hosted at Singapore University of Technology and Design (SUTD) collaborated in testing the OFH interface of an O-DU from Synergy Design Technology\footnote{\url{https://www.synergydesigntek.com/}}. In collaboration with Keysight and Calnex (test and measurement solutions), the two OTICs have evaluated several test cases for the CUSM-Plane (i.e., control, user-data, synchronization, management planes).

While existing literature and industry efforts have made progress in defining test specifications and methodologies for O-RAN components, there is a notable gap in comprehensive studies that analyze the challenges and best practices for conducting consistent O-DU testing in diverse setups. This gap is particularly evident in the context of conformance testing, which is crucial for ensuring compliance with O-RAN specifications.
%
%
To bridge this gap, our paper presents a study on testing O-DU implementations across multiple O-RAN test and integration labs. We focus on the following key contributions.

\begin{itemize}
\item We present the O-DU testing setups and procedures across two geographically distributed OTIC labs, highlighting the complexities involved in establishing a consistent test environment.
\item We identify and examine key factors affecting test repeatability, such as the test equipment, timing configuration, and the virtualization technologies used. 
\item We discuss the challenges encountered during O-DU testing, and propose follow-up work needed from O-RAN ALLIANCE and the whole O-RAN community (including OTICs, vendors) to address them. 
\item We propose best practices and review lessons learned to enhance the consistency and reliability of O-DU conformance and performance testing across different lab environments.
\end{itemize}

\section{Testing Setup across OTIC Labs}
\label{sec:setup}

In this section, we will briefly describe the Device-under-test (DUT) and the standard setup to measure CUSM-Plane for such DUT.
We then present testing setups deployed in two OTICs at SUTD and NEU, pointing out differences and their possible influence on the result.

\subsection{Device-under-test (DUT)}


The O-DU is a key element in the 5G radio access network (RAN). It hosts RLC/MAC/High-PHY layers and supports lower protocol stack layers, including baseband processing and RF functions. 
The O-DU is controlled by the O-RAN Central Unit (O-CU), and an O-CU can be linked to multiple O-DUs.



The Genevisio’s PAC series inline accelerator card \cite{Genevisio_DU_card} is an ASIC-based PCIe DU solution which leverages NXP's Layerscape Access family of programmable baseband processors (LA-12 series + LX-2 series) to deliver multiple 25GbE eCPRI interfaces and maximum support for four 4T4R 100MHz RUs.
Synergy offers software designed for the PAC series inline accelerator card, which fulfils the role of an O-DU within a 5G RAN system. The Synergy O-DU card is shown as the O-DU part of the DUT in Fig.~\ref{fig:cum-plane-test-setup}.

\subsection{Testing Scenarios}
\label{subsec:TestScenarios}


\begin{figure}[!t]
    \centering
    \includegraphics[width=0.6\linewidth]{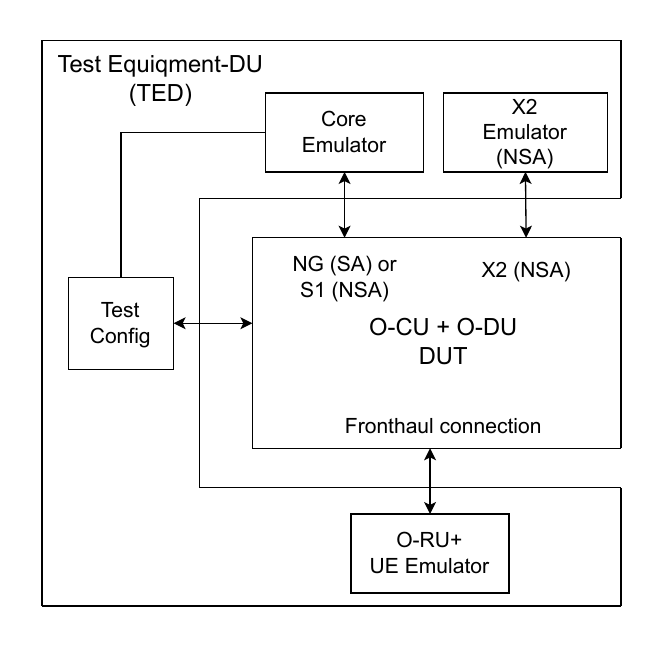}
        \vspace{-.45cm}
    \caption{Combined O-CU and O-DU (from Synergy) as the DUT}
    \label{fig:Combined-OCDU-as-the-DUT}
    \vspace{-.45cm}
\end{figure}

As stated in the O-RAN Working Group 4 (WG4) conformance test specification~\cite{oran_conf_test}, two test configurations are available for the O-DU as the Device-Under-Test (DUT): the Stand-alone O-DU as the DUT, and the Combined O-CU+O-DU (from the same vendor) as the DUT. In our current setup, we have opted for the latter configuration as illustrated in Fig.~\ref{fig:Combined-OCDU-as-the-DUT}. In particular, our test setup --- called the {\it Test Equipment-DU (TED)} in the following --- encompasses the combined O-CU and O-DU as the DUT through the NG (SA) interfaces toward the Core Emulator, the OFH interface toward the O-RU+UE Emulator (i.e., RuSIM), and the O1 interface toward the test configuration entity.

\subsection{Configuration of CUSM-Plane Tests}
\label{subsec:ConfigurationCUSM}

\begin{figure}[!t]
    \centering
    \includegraphics[width=1.0\linewidth]{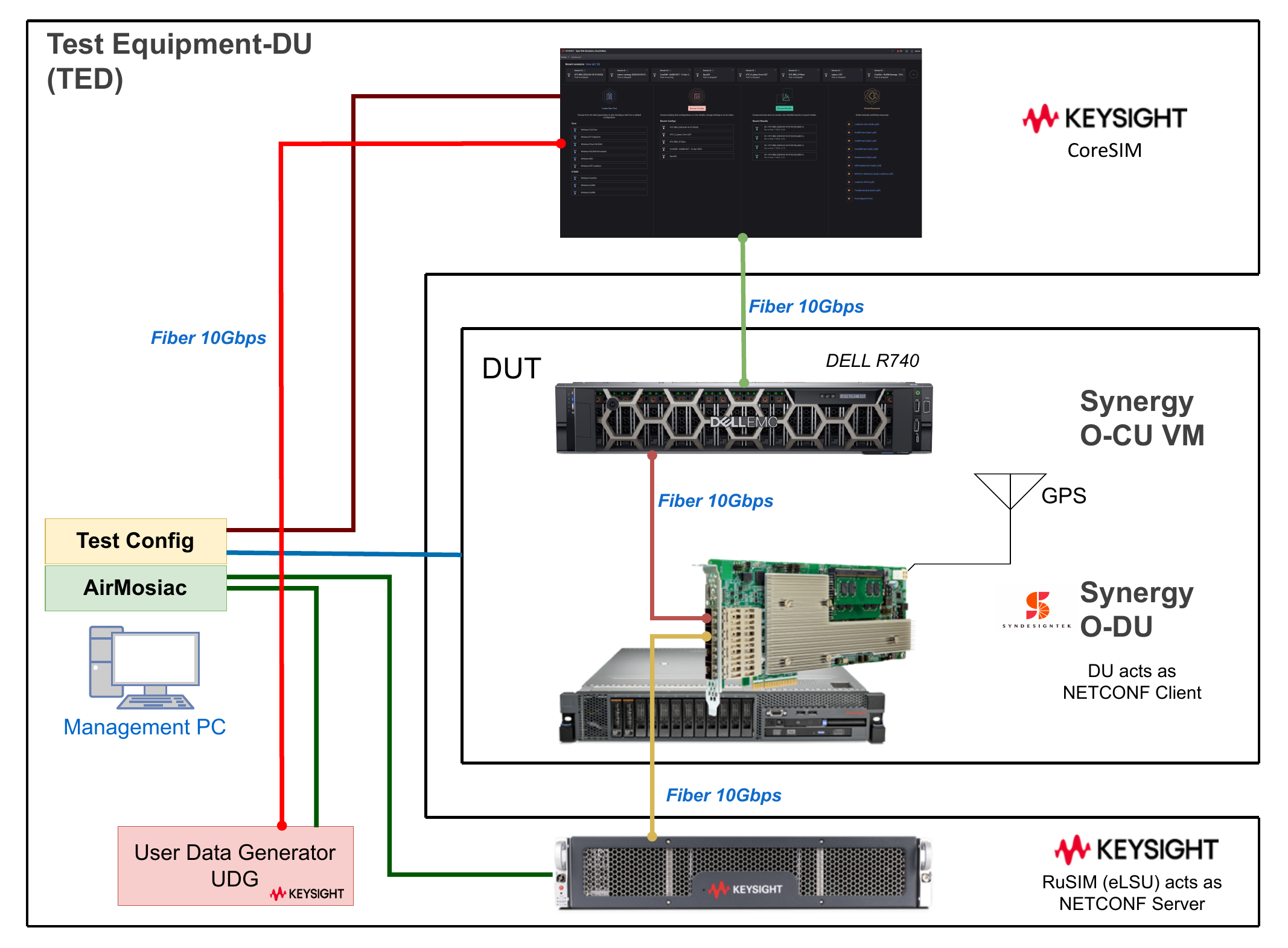}
    \caption{O-DU CUM-Plane Test setup at Singapore OTIC}
    \label{fig:cum-plane-test-setup}
    \vspace{-.45cm}
\end{figure}

\textbf{M-Plane} uses the NETCONF network element management protocol and the YANG data modeling language. Within the NETCONF framework, YANG defines the structure and constraints of the data exchanged between the client and server. YANG models describe network elements' configuration, state, and operational data, providing standardized data definitions across vendors and devices. In the NETCONF framework, a client and server model facilitates information exchange and execution of remote procedure calls (RPCs). Each NETCONF transaction involves a bidirectional exchange between the client and server to ensure seamless communication. In our testing environment, the O-DU acts as a NETCONF client, while the RuSIM functions as a NETCONF server, as shown in Fig. \ref{fig:cum-plane-test-setup}. For the first test case in M-Plane, Section 3.1.1 in Conformance Test Specification \cite{oran_conf_test}, packets exchanged between the NETCONF client and the NETCONF server are captured and analyzed using Wireshark. For all remaining M-Plane test cases, the analysis relies solely on logs obtained from either side. This is because of the encryption of messages exchanged between the client and server.


\begin{figure}[h!]
    \centering
    \includegraphics[width=1.0\linewidth]{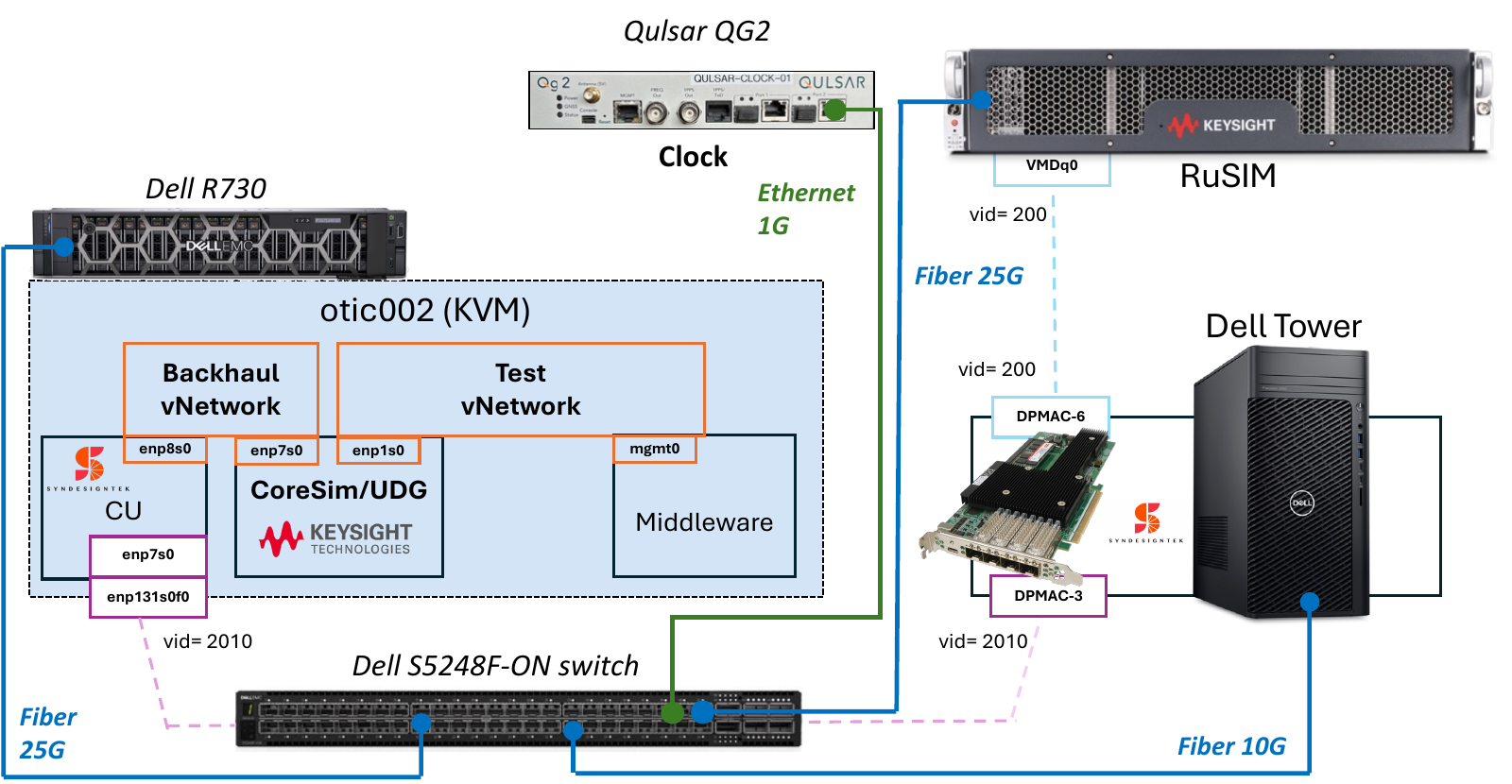}
    \caption{O-DU UC-Plane Test Setup in NEU}
    \label{fig:cum-plane-test-setup_NEU}
        \vspace{-.45cm}
\end{figure}

\textbf{UC-Plane}
Fig.~\ref{fig:cum-plane-test-setup} and Fig.~\ref{fig:cum-plane-test-setup_NEU} show the topology setup at Singapore OTIC and OTIC at NEU for UC-Plane testing respectively.
Both OTICs utilized tools provided by Keysight as TED for testing UC-Plane.
The CoreSIM, as shown on the top of Fig.~\ref{fig:cum-plane-test-setup}, consists of a graphical interface for management and multiple CoreSIM agents. Each CoreSIM agent is deployed on a dedicated virtual machine to support testing multiple DUTs in parallel.
CoreSIM presents support for three distinct network stacks: Linux Stack, Keysight's IxStack
over Raw Socket, and IxStack over DPDK.
While conducting the test cases, it is observed in Singapore OTIC that IxStack over DPDK demonstrates the highest level of throughput performance.

At the bottom of Fig.~\ref{fig:cum-plane-test-setup}, Keysight eLSU (Ethernet Line Server Unit) is a versatile device capable of functioning as either a RuSIM or UeSIM. In its RuSIM capacity, it emulates RU and UEs' functionalities. Keysight Airmosaic software controls RuSIM, enabling the creation of multiple UEs with customized configurations. Proper alignment between the UEs and CoreSIM configurations is imperative to enable UE registration with the network. 
The User Data Generator (UDG) software (i.e., red box in Fig.~\ref{fig:cum-plane-test-setup}) is utilized to generate user traffic and is operated under the supervision of Airmosaic. 


\begin{figure}[h!]
    \centering
    \includegraphics[width=0.85\linewidth]{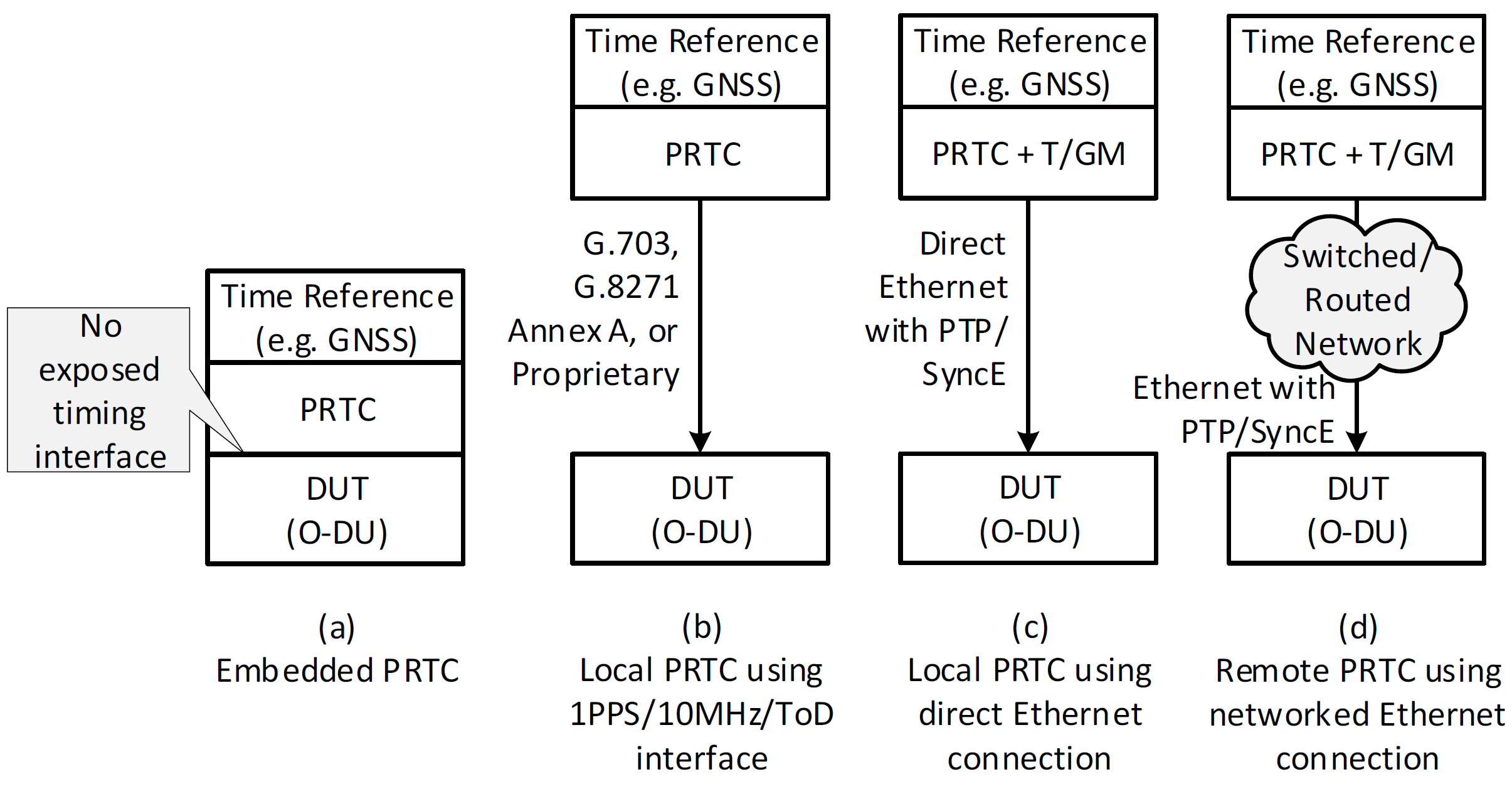}
    \vspace{-.3cm}
    \caption{Methods to connect PRTC to O-DU \cite{oran_conf_test}}
    \label{fig:methods_to_connect_prtc_to_odu}
    \vspace{-.3cm}
\end{figure}

\textbf{S-Plane}
For S-Plane testing, the ORAN conformance test \cite{oran_conf_test} recommends four specific connection types for both local and remote PRTC connections, with detailed specifications for the PRTC to O-DU connection. In our particular setup, we tested connection types (c) and (d), as depicted in Fig. \ref{fig:methods_to_connect_prtc_to_odu}. In this configuration, the Primary Reference Time Clock (PRTC) is interfaced with the O-DU through either a direct Ethernet connection or a networked Ethernet connection.

\begin{figure}[!t]
    \centering
    \includegraphics[width=0.8\linewidth]{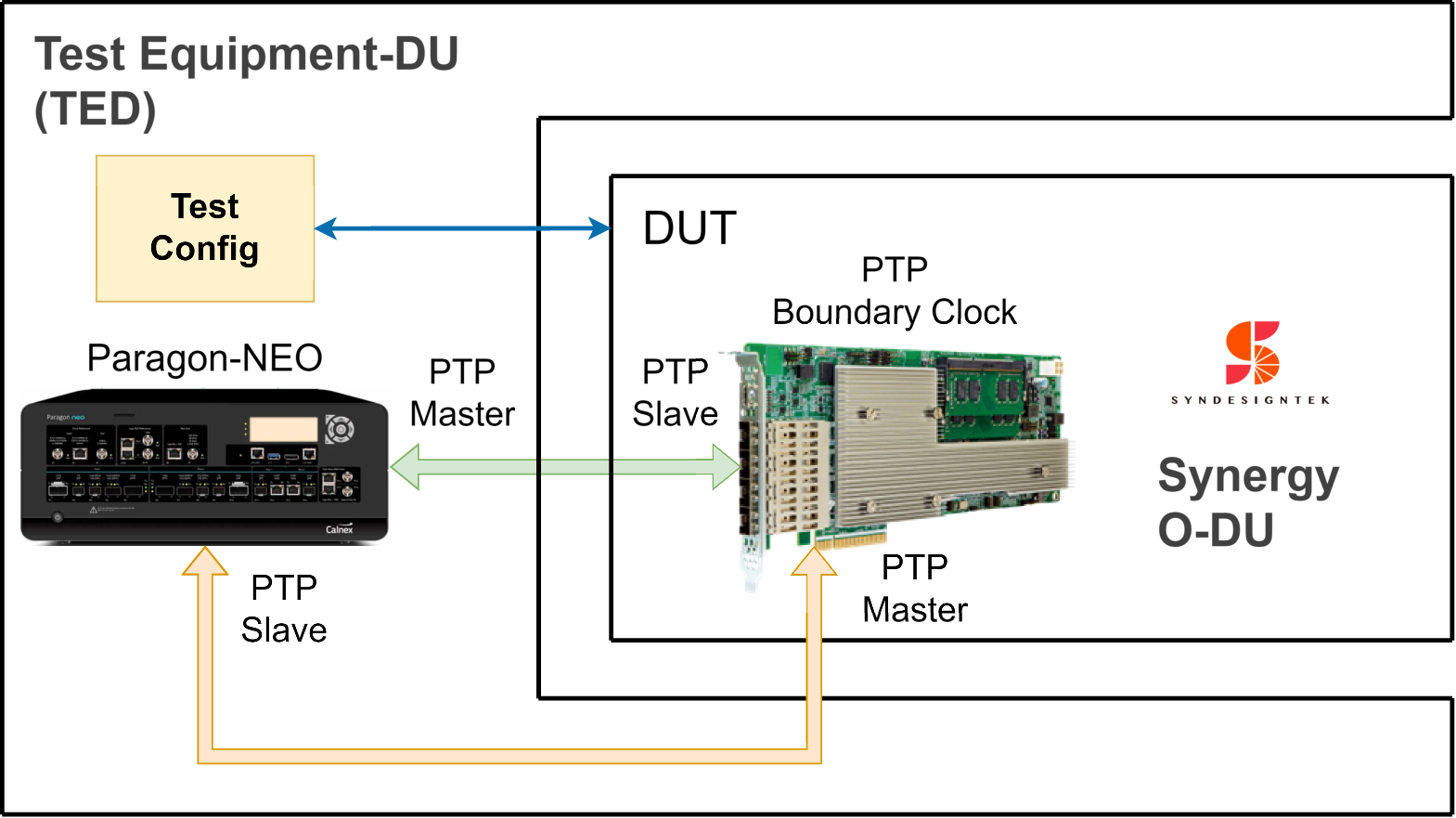}
    \caption{O-DU S-Plane Test setup at Singapore OTIC}
    \label{fig:s-plane-test-setup}
    \vspace{-.45cm}
\end{figure}

In the Singapore OTIC setup, the Paragon-Neo from Calnex\cite{Calnex_ParagonNeo} is used for S-Plane testing as illustrated in Fig.~\ref{fig:s-plane-test-setup}. This system is capable of emulating Precision Time Protocol (PTP) Master and Slave clocks to enhance the precision and consistency of PTP tests while offering specific test modes tailored to various DUTs, along with automatic test selection for compliance with O-RAN standards. Moreover, Paragon-Neo can replicate a switched/routed network, facilitating direct connection to the O-DU for testing both (c) and (d) setups.

In the NEU setup, the Qulsar QG-2 is employed for S-Plane testing as shown in Fig.~\ref{fig:cum-plane-test-setup_NEU}. The QG-2 functions as a PTP Grandmaster, providing a highly accurate time source for synchronization purposes. This setup allows for the precise evaluation of the S-Plane's (c) case, which involves verifying the synchronization accuracy and stability of the O-DU.

\subsection{Differences between the Two OTICs' Setups}
\label{subsec:DifferencesSetups}
For UC-Plane test, Singapore OTIC used LLS-C1 topology with an \textit{embedded} PRTC time source at O-DU to achieve synchronization between O-DU and O-RU. 
In the NEU setup, the initial selection of the LLS-C3 topology was made for ease of setup. However, due to instability, it was subsequently replaced with the LLS-C1 topology with a \textit{remote} PRTC time source (as shown in option (d) in Fig.~\ref{fig:methods_to_connect_prtc_to_odu}). In this setup, the PRTC/T-GM being utilized is the Qulsar QG-2, which is functioning as a PTP grandmaster.

The deployment of CoreSIM agents and O-CU at the SUTD involves the utilization of VMware ESXi virtualization technology, which operates on two different Dell PowerEdge R740. Meanwhile, at NEU, the chosen virtualization technology is a Kernel-based Virtual Machine (KVM) where the O-CU and CoreSIM agents are deployed in separate virtual machines, both running on a single Dell R730 rack server.

For UDG, there is a slight difference between SUTD and NEU setup. SUTD hosts UDG software inside the eLSU server, while NEU hosts the UDG software on the same VM running the CoreSim Agent. It is observed that this discrepancy does not affect the test outcome.


\subsection{Test Execution}

Prior to the Spring 2024 PlugFest, the Singapore OTIC had used and obtained some experience with Synergy O-DU for a few months. 
During PlugFest, Singapore OTIC spent two months conducting the majority of O-DU conformance test cases and some performance test cases. Throughout this process, Singapore OTIC worked closely with Synergy, Keysight, and Calnex to align the configuration between DUT and TED. OTIC at NEU, on the other hand, only had 2 weeks to set up and test the O-DU due to PlugFest-specific timelines.

To ensure consistent and repeatable results, the two OTICs agreed to execute the same list of test cases. In addition, we also shared 
detailed preparation in the following aspects:


\begin{itemize}


\item Design detailed test setups with accompanying diagrams to provide a clear overview of the infrastructure.

\item Align software versions in the TED to prevent compatibility issues and ensure accurate test results.

\end{itemize}


\section{Test Results}
\label{sec:results}

\textbf{Conformance Test:} A majority of the M-Plane test cases have been successfully carried out at the Singapore OTIC. However, due to time constraints, these M-Plane test cases were not executed at NEU. For S-Plane and UC-Plane, as shown in  table~\ref{tab:DU_results_consistent}, the O-DU conformance test results are consistent and repeatable in the two OTICs in NEU and Singapore. There are some notes for test cases 3.3.5 and 3.4.4.4.1. In 3.3.5, the PTP sync-state shown by O-DU may be a different text but has the same meaning as the state defined in the conformance test specification. For instance, the O-DU may show its sync-state ``Not\_Locked'' instead of ``FREERUN''. In test case 3.4.4.4.1, while other parameters work as defined in IOT profiles \cite{o_ran_iot_specification},  DU needs to adjust its transmit time of uplink control packets, which is outside the standard IOT profile, to make these packets arrive on time.

\begin{table}[!t]
\begin{center}
   \caption{Synergy O-DU Conformance Test results (see additional notes in Section~\ref{sec:results} for Pass* results) 
   }
   \vspace{-.15cm}
   \footnotesize
\begin{tabular} 
{p{1.1cm} | p{4cm} | p{1cm} | p{1cm}}
\toprule
  Test case &  Description & NEU & Singapore  \\ 
  \midrule
  3.3.5 & Func. test of O-DU Sync. from ITU-T G.8275.1 profile PRTC/T-GM (LLS-C1/C2/C3/C4) & Pass* & Pass* \\ 
  \midrule
  3.4.4.1.1 & UC-Plane O-DU Scenario Class NR testing Generic (NRG) 
 & Pass & Pass \\ 
  \midrule
  3.4.4.3.1 & Static Format Fixed-Point (FP) Uncompressed
 & Pass & Pass \\
  \midrule
  3.4.4.3.2 & UC-Plane O-DU Scenario Class Compression (CMP) Static Format Block Floating Point
 & Pass & Pass \\
  \midrule
  3.4.4.4.1 & Delay Management On-time arrival
 & Pass* & Pass* \\
\bottomrule
\end{tabular}
\label{tab:DU_results_consistent}
\end{center}
\vspace{-3mm}
\end{table}

  
  
  


\begin{table}[!t]
\begin{center}
\caption{TIFG E2E Performance, Peak Throughput (Mbps)  
\vspace{-.15cm}
\cite{o_ran_end_to_end_test_specification}}
   \footnotesize
\begin{tabular}{c | l | c | c}
  \toprule
  Test case & Description & NEU & Singapore \\ 
  \midrule
  6.2 & Downlink peak throughput
 & 900 & \textbf{1550} \\ 
  6.4 & Uplink peak throughput
 & 175 & \textbf{525} \\ 
  \bottomrule
\end{tabular}
\label{tab:tifg_e2e_performance}
\end{center}
\vspace{-3mm}
\end{table}

\textbf{Performance Test:}
In the context of the performance test, the difference in deployment methods of O-CU in the two OTICs has led to divergent results. O-CU in SUTD is deployed on VMware ESXi virtualization technology running on DELL R740, while the deployment at NEU is based on KVM deployed on DELL R730.

\textbf{Issues Encountered and Resolutions:}
In our initial attempts, the O-DU tests encountered several failure scenarios, especially in M-Plane. For instance, Synergy's DU initially failed to parse some YANG models received due to its implementation, which led to later failures in parsing messages sent from RuSIM. On the other hand, RuSIM's default setting leaves some fields empty, which are defined as non-empty fields in the O-RAN model specification. After the issues were identified in our tests, the DUT and TED vendors coordinated and updated their software to fix the problem. In another instance, we observe the `PtpClockIdentity' on S-Plane, which is a custom field in Paragon-Neo, requires users to fill in the value. This value should be set according to IEEE EUI-64 assigned numbers as indicated in RFC 8173 \cite{rfc_ptp_mib}. Under the default value used by the Paragon-Neo, Synergy's DU was unable to synchronize with Paragon-Neo.

\section{Challenges and Lessons Learnt}
\label{sec:challenges}

In this section, we describe the challenges and lessons learned and propose suggestions for achieving consistent and repeatable results across OTICs.

\subsection{Challenges}

\textbf{Identifying Causes of Test Failures:} As shared in previous Section, multiple factors can contribute to test failures or inconsistency observed during testing. In some cases, determining the precise root cause is a considerable challenge, potentially stemming from (i) the DUT, (ii) the test \& measurement tools, (iii) errata in the specifications, or (iv) erroneous tester operations. Notably, convincing to the DUT vendor that the issue originated from their device often necessitated a considerable amount of time. On the other hand, TED-related issues require collaboration with the TED developer to resolve them. Problems can also be from the specification. While conducting the tests, an error was identified in the imported O-RAN YANG model. Although the IETF has rectified the errata \cite{rfc_errata} and the O-RAN ALLIANCE has outlined a plan to address the errata present in the imported models, at the time of testing, the official O-RAN release had not been updated to reflect the correction.

\textbf{Repeatability of Testing Results:} The outcome of a given test case depends on the stability of both the DUT and the TED. While a test case may demonstrate success following a DUT restart in a pristine state, failure is likely if the DUT has been operational for an extended duration or if the DUT experiences a malfunction during the test. Consequently, evaluating the pass or fail status of a conformance test proves to be a challenging task.

\textbf{Ambiguity in Features Supported by DUT:} The DUT vendor and OTIC may have different understandings regarding the features needed to pass test cases in O-RAN test specification. For instance, initially, Synergy's DU only supports embedded PRTC, which is connection type (a) in figure \ref{fig:methods_to_connect_prtc_to_odu}, but was reported to support test case 3.3.5 in the conformance test specification~\cite{oran_conf_test} where O-DU is synchronized using the Ethernet interface. We suggest OTICs to provide a clear explanation of the test case and engage in a discussion with the vendors to verify the supported features. 

\textbf{Technical Support and Coordination across Different Time Zones:} The testing procedures were conducted by two OTICs located on different continents with 12 hours of time difference. Subsequently, the exchange of information is impeded not only by geographical separation but also by time zone disparities. In urgent situations, direct setup and configuration assistance were typically extended during nighttime (for one party) to facilitate prompt resolution. 

\textbf{Difficulties in Automating O-DU Testing:} In O-DU testing, the automation of O-DU implementation relies on vendor-specific interfaces of the O-DU. Within the M-Plane, the O-DU functions as a NETCONF client, and the manner to support user interaction with this client is decided by the O-DU vendor. In the case of Synergy O-DU, user interaction with the client occurs via a command line interface. As for the UC-Plane, configuration changes are also contingent on the vendor. These factors contribute to making the implementation of automation testing for the O-DU more complex compared to the O-RU.

\subsection{Lessons Learned}


\textbf{Virtualization and Compute Technology:} The selection of virtualization technology can have a significant impact on system throughput as shown in table \ref{tab:tifg_e2e_performance}. With the default configuration in both OTIC labs, it has been established that the utilization of VMware ESXi virtualization technology on the newer Dell server yields notably higher downlink throughput than KVM on the older Dell machine. After an exhaustive debugging process, it was ascertained that the predominant packet drops occur at O-CU running on KVM, which is attributed to its inability to process a high throughput of traffic effectively. To achieve consistent testing results, it is advisable to utilize identical virtualization technology and configurations for CU in all OTICs, or to fine-tune the configuration on both virtualization solutions to achieve similar performance.


\textbf{Packet Size and Burstiness:} In the throughput test cases, it is evident that the size of each packet significantly influences the maximum throughput. A traffic generator has the capability to attain equivalent throughput using two methods: by transmitting small packets with a high frequency or by transmitting large packets less frequently. It has been observed that the overall system throughput can achieve the highest value when using the transmission of large packets. This is attributed to the faster nature of copying large packets in comparison with processing a larger number of smaller packets. For the purpose of achieving consistent results, it is advisable to measure throughput in conjunction with the corresponding packet size and burstiness.

\textbf{On-time Arrival Problem:} Keysight RuSIM has the capability to synchronize with the DU solely through the reception of UC-Plane packets without S-Plane, a feature specific to Keysight. To validate the time-related test cases, it is essential to configure RuSIM to utilize S-Plane for synchronization. Moreover, prior to commencing testing,
as described in section 11.7.2 of \cite{oran_cus_plane_spec},
we must ensure that the configuration for $\alpha$ and $\beta$ are identical between O-DU and RuSIM. One more factor that can affect the on-time arrival test cases is the PTP timescale. While RuSIM can be configured to operate based on either Coordinated Universal Time (UTC) or International Atomic Time (TAI) references, it operates in UTC by default. Synergy O-DU, on the other hand, explicitly utilizes the TAI. Currently, there is a discrepancy of 37 seconds \cite{timeanddate} between UTC and TAI. It will result in an inaccurate count of on-time arrival packets on RU if there is a mismatch in the configuration of the PTP timescale between DU and RU.
%


\textbf{Communication between two OTICs:} is crucial to avoid misunderstandings about pass or fail criteria. Personnel from two OTICs should share their testing experiences to ensure a clear understanding of testing scenarios and potential issues.

\textbf{Test Automation for Open RAN Systems:} Utilizing test automation in Open RAN systems is important for ensuring predictability and consistency of outcomes. Currently, the absence of an automation tool for testing O-DU necessitates manual testing. 
The implementation of automation can reduce human error and save time and resources in cases where repetitive test runs are required.

\section{Conclusions}
\label{sec:conclusions}


In this paper, we presented a joint study by the Asia \& Pacific OTIC in Singapore and the North America OTIC in the Boston Area at Northeastern University to test OFH CUSM-Plane of the Synergy O-DU. We highlighted differences between the two setups for testing the different planes. We analyzed which differences could affect the testing results. Finally, we shared challenges and learned lessons during the PlugFest, which was jointly hosted by the two OTIC Labs located on different continents. We emphasized the importance of configuration alignment between DUT and TED on conformance tests. For the performance test, we indicated the influence of virtualization technology deployed in different platforms on the test result and the impact of packet size and burstiness on throughput measurement. These findings could help the O-RAN ALLIANCE define more specific test cases to archive consistent and repeatable testing results across OTICs. In our future work, we will conduct the remaining O-DU conformance tests specified in~\cite{oran_conf_test} and enhance the automation level of conducting the tests.

\section*{Acknowledgment}
This research was partially supported by the National Research Foundation, Singapore and Infocomm Media Development Authority (IMDA) under its Future Communications Research \& Development Programme, by the Massachusetts Technology Collaborative under award 22563, and by the National Telecommunications and Information Administration (NTIA)'s Public Wireless Supply Chain Innovation Fund (PWSCIF) under Award No. 25-60-IF054. Any opinions, findings and conclusions expressed in this material are those of the author(s) and do not reflect the views of the funding agencies. 

\bibliographystyle{IEEEtran}
\bibliography{refs}

\end{document}